\documentclass[aps,pra,superscriptaddress,twocolumn]{revtex4-1}
\usepackage{dcolumn}
\usepackage[utf8]{inputenc}
\usepackage{amsmath}
\usepackage{amsfonts}
\usepackage{amssymb}
\usepackage{graphicx}
\usepackage{hyperref}
\usepackage{xcolor}

\newcommand{\ket}[1]{\ensuremath{\left|{#1}\right\rangle}}

\begin{document}
\title{Engineering entangled photons for transmission in ring-core optical fibers} 
\author{G. Ca\~{n}as}
\affiliation{Departamento de F\'isica, Universidad del B\'io-B\'io, Collao 1202, 5-C Concepci\'on, Chile}
\author{E. S. G\'omez}
\affiliation{Departamento de F\'{\i}sica, Universidad de Concepci\'on, 160-C Concepci\'on, Chile}
\affiliation{Millennium Institute for Research in Optics, Universidad de Concepci\'on, 160-C Concepci\'on, Chile}
\author{E. Baradit}
\affiliation{Departamento de F\'isica, Universidad del B\'io-B\'io, Collao 1202, 5-C Concepci\'on, Chile}
\author{G. Lima}
\affiliation{Departamento de F\'{\i}sica, Universidad de Concepci\'on, 160-C Concepci\'on, Chile}
\affiliation{Millennium Institute for Research in Optics, Universidad de Concepci\'on, 160-C Concepci\'on, Chile}
\author{S. P. Walborn}
\affiliation{Departamento de F\'{\i}sica, Universidad de Concepci\'on, 160-C Concepci\'on, Chile}
\affiliation{Millennium Institute for Research in Optics, Universidad de Concepci\'on, 160-C Concepci\'on, Chile}

\begin{abstract}
The capacity of optical communication channels can be increased by space division multiplexing in structured optical fibers. Radial core optical fibers allows for the propagation of twisted light--eigenmodes of orbital angular momentum, which have attracted considerable attention for high-dimensional quantum information.   Here we study the generation of entangled photons {that are tailor-made for} coupling into ring core optical fibers. We show that the coupling of photon pairs produced by parametric down-conversion can be increased by close to a factor of 3 by pumping the non-linear crystal with a perfect vortex mode with orbital angular momentum $\ell$, rather than a gaussian mode.    Moreover, the two-photon orbital angular momentum spectrum has a nearly constant shape.  This  provides an interesting scenario for quantum state engineering, as pumping the crystal with a superposition of perfect vortex modes can be used in conjunction with the mode filtering properties of the ring core fiber to produce simple and interesting quantum states. 

\end{abstract}
\maketitle

\section{Introduction}
           
\par
Distribution of photonic entangled states is a cornerstone of future quantum networks.  Most likely, this will need to be realized within the same optical infrastructure as standard telecommunications networks.  Recent developments in optical fiber technology have resulted in novel fiber core structures, which allow for the propagation of multiple spatial modes. These fibers are expected to play an important role in increasing the transmission capacity of future telecommunications networks through space division multiplexing (SDM) \citep{richardson13}.   Examples of SDM fiber candidates include multi-mode fibers \citep{Sillard, Rademacher}, multi-core fibers \citep{Saitoh}, and ring core fibers \citep{RCF}, among others.  In the quantum regime, SDM technology has attractive features. The multiple spatial modes are a straightforward way to increase the dimension of quantum systems, which has several advantages in quantum key distribution (QKD) \citep{bourennane01,collins02,cerf02,walborn06a,huber13,Canas17}, and have shown to be more resistant to some types of noise \citep{zhu21}. {Additional applications can be found in a recent review \cite{erhard20}.}    In addition to providing multiple channels, it is expected that these fibers will offer more phase stability, when compared to superposition states of several modes propagating in independent fibers \citep{ding17,xavier20}. 
\par
Ring-core fibers (RCFs) allow for the propagation of orbital angular momentum (OAM) eigenmodes \citep{bozinovic13,Nejad,Gregg15,Ramachandran15, Gregg16,RCF_Wang,Zhu:18,Bacco2019,Zhang:20,rojas21}, which have attracted considerable attention as they allow for the encoding of high-dimensional quantum information \citep{bavaresco18,erhard18,forbes19}.   
  %To date, the principal source of entangled photons has been spontaneous parametric down-conversion (SPDC), where the photons can be entangled in polarization \citep{ou88,shih88,kwiat95}, linear momentum \citep{rarity90b}, time-bins \citep{tapster94}, transverse position-momentum \citep{dangelo04,howell04,tasca08}, and orbital angular momentum \citep{mair01,walborn04a,straupe11}. 
   To date, the principal source of entangled photons has been spontaneous parametric down-conversion (SPDC). A beautiful and useful characteristic of SPDC is that the two photon spatial state can be engineered by manipulating the pump beam \citep{monken98a,torres03a,Miatto11,Yao11,Kovlakov17,Liu18,Kovlakov18}. This has led to  the production of quantum states with interesting properties \citep{walborn07c,nogueira04,hernandez16,lopez19,baghdasaryan20}.   The entanglement properties of these states are determined by the mode decomposition of the two-photon state, in which one has the freedom to choose between quite a few families of transverse modes. Of particular interest are those decompositions onto OAM eigen-modes such as Laguerre-Gauss \citep{arnaut01,franke-arnold02,torres03b,walborn04a} or Bessel-Gauss \citep{mclaren12,mclaren13,anwar2021size},  which comprise a set of Schmidt modes of the two-photon state \citep{law04,straupe11}. 
   %%%%%%%%%%%%%%%%%%%
\begin{figure}
\begin{center}
\includegraphics[width=7cm]{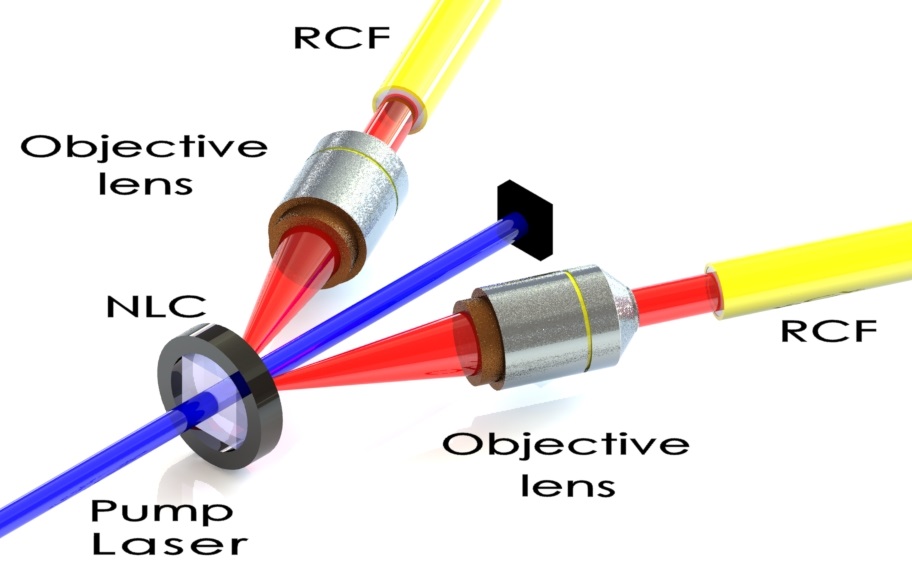}
\caption{Sketch of the basic idea. SPDC at a thin non-linear crystal (NLC) produces down-converted photon pairs, which are then coupled into ring-core fibers. The angle between the down-converted beams and the pump beam is exaggerated in the figure for visual clarity.  In the calculations, it is assumed to be small enough so that the setup can be treated as co-linear. The optical systems are such that the image plane of the crystal coincides with the entrance face of the fibers, with appropriate magnification factors.} 
\label{fig:basicidea}
\end{center}
\end{figure}
%%%%%%%%%%%%%%%%%%%
\par
In this paper we study the coupling of down-converted photons into RCFs and the two-photon state that is produced, as sketched in Fig. \ref{fig:basicidea}.  We consider the decomposition of the two-photon state in terms of perfect vortex (PV) modes, which can have near-perfect fidelity with the eigen-modes of RCFs \citep{rojas21}.  We show that by pumping the down-conversion crystal with a PV pump beam, the amplitude of the most relevant down-converted PV modes can be increased while maintaining a high degree of entanglement. While the two-photon OAM mode spectrum is wider for a gaussian pump beam, leading to larger Schmidt numbers, this increase in entanglement is irrelevant when coupling into RCFs or similar optical devices, as only a finite set of lower-order PV modes excite the fiber eigen-modes.  In addition, the shape of the mode spectrum is nearly independent of the OAM of the PV pump beam. The combination of the two-photon state engineering using PV pump beams and the mode filtering provided by the RCF can be a powerful tool, providing a simple method to generate interesting two-photon states, {some examples of which are} discussed in section \ref{Sec:Apps}.            

\section{Perfect Vortex Beams and Ring Core Fibers}

%%%%%%%%%%%%%%%%%%%
\begin{figure}
\begin{center}
\includegraphics[width=8cm]{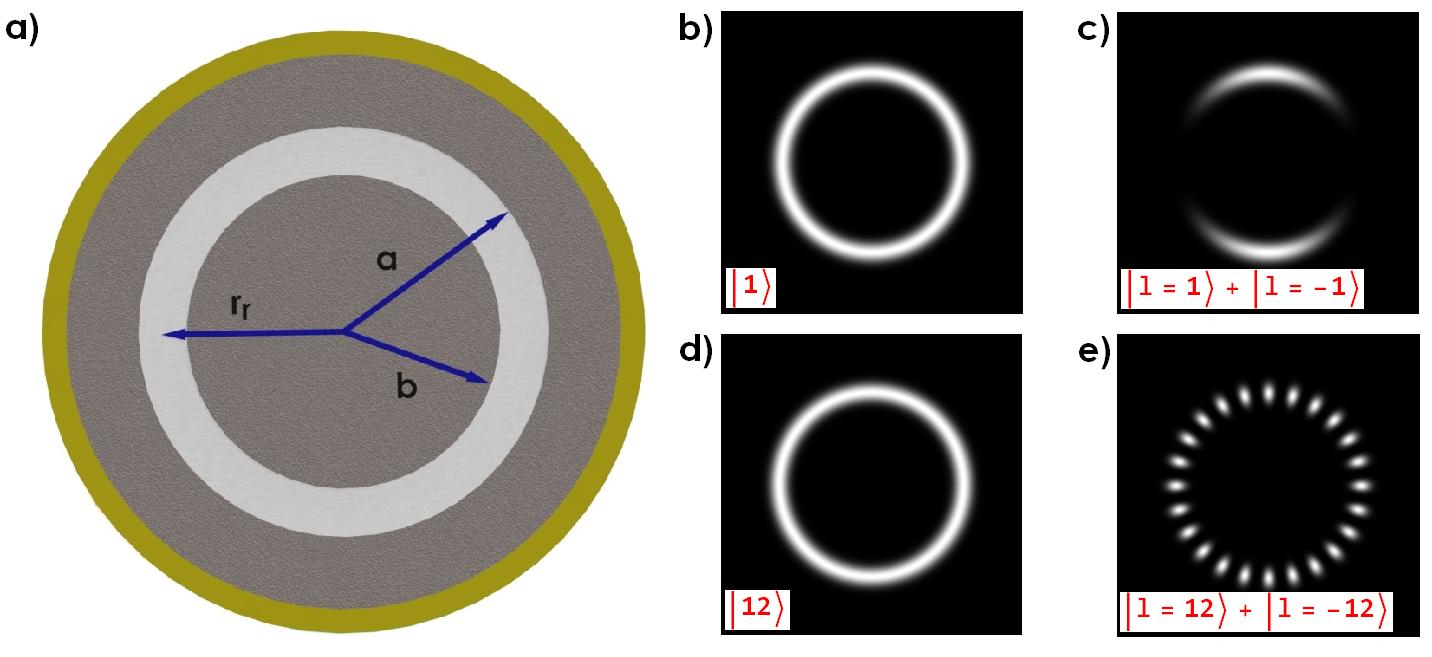}
\caption{a) Illustration of a Ring-Core Fiber (RCF).  b), d) Profiles of perfect vortex modes and c), e) superpositions  of perfect vortex modes.} 
\label{fig:rcf}
\end{center}
\end{figure}
%%%%%%%%%%%%%%%%%%%
An illustration of a RCF is shown in Fig. \ref{fig:rcf}.  It is described by a ring-shaped core, with interior radius $b$ and exterior radius $a$.   A set of eigenmodes of the RCF (circularly symmetric LP modes) have an azimuthal phase dependence $e^{i \ell \phi}$, and thus carry OAM \citep{allen92}.  In Ref. \citep{rojas21} it was shown that an example of a commercially available RCF supports 13 LP modes with OAM  $\ell=0,\pm 1, \pm 2, \dots \pm 6$.  Moreover, depending on the fiber properties, they can have near perfect ($\sim 0.995$) overlap with the so-called PV modes with the same value of $\ell$. PV  modes are the Fourier transform of  Bessel-Gaussian beams, carrying OAM with eigenvalue $\ell$. They are given by \citep{vaity15} 
\begin{equation}
\label{eq:pvb}
  \mathcal{U}_{\ell}(\boldsymbol{r}) =N \exp(i\ell\phi) u_\ell(r),
\end{equation}
where $N$ is a normalization constant and the radial component is given by
\begin{equation}\label{eq:pvb3}
  u_{\ell}(r) =   \exp\left(-\frac{(r^2+r_r^2)}{w_0^2}\right) I_\ell \left ( \frac{2 r r_r}{w_0^2} \right), 
\end{equation}
with $I_\ell$ the modified Bessel function of the first kind. The parameters  $r_r$ and $w_0$ are the ring radius and the Gaussian beam waist at the focus.  As shown in Ref. \citep{vaity15}, with certain parameter relations, the radial function can be approximated by  
\begin{equation}\label{eq:pvb2}
  u_{\ell}(r) \approx    u(r) = \exp\left(-\frac{(r-r_r)^2}{w_0^2}\right),
\end{equation}
which presents the interesting property that it is independent of $\ell$.    Numerical evaluation of the overlap between normalized versions of  \eqref{eq:pvb3} and \eqref{eq:pvb2} is near unity when $\ell \lesssim 3 r_r/w_0$. The ring-shape with constant radius makes these modes attractive for coupling into RCFs \citep{Brunet:14,P_Vaity,vaity15,rojas21}. Thus, entangled photons in perfect vortex modes are an interesting candidate for distribution of entanglement in RCFs.   

%Fig. \ref{fig:overlap} shows a plot of the absolute value of the overlap of  PV modes \eqref{eq:pvb} with the same $\ell$ and radial functions given by \eqref{eq:pvb3} and \eqref{eq:pvb2}. As the figure shows, for a wide range of values of {$r_r/w_0$} that depends upon $\ell$, the PVBs are well described by the radial function \eqref{eq:pvb2}. 

%%%%%%%%%%%%%%%%%%%
%\begin{figure}
%\includegraphics[width=7cm]{Overlap-PVB-PVB_exact}
%\caption{Absolute value of the verlap between PV modes described by \eqref{eq:pvb} with radial functions given by \eqref{eq:pvb3} and \eqref{eq:pvb2}. } 
%\label{fig:overlap}
%\end{figure}
%%%%%%%%%%%%%%%%%%%
  
\par

%The achievable coupling efficiency of PVBs into a commercially available RCF has been shown to be near unity ($\sim 0.995$), when the ring radius $r_r$ and width $w_0$  are given by $0.8??? a$ and $0.475 a$, respectively \citep{rojas21}.
\section{Spontaneous Parametric Down-Conversion with PV Modes}
\label{Sec:SME_SPDC}
The two-photon state produced from SPDC using a continuous-wave, monochromatic pump beam incident on a thin non-linear crystal, is given by \citep{hong85,monken98a,walborn10,Schneeloch16}
\begin{equation}
\ket{\psi}= \iint  d\boldsymbol{q}_1 d\boldsymbol{q}_2 \psi (\boldsymbol{q}_1,\boldsymbol{q}_2) \ket{\boldsymbol{q}_1} \ket{\boldsymbol{q}_2}, 
\label{eq:cspdc}
\end{equation}
where $\boldsymbol{q}_l$ ($l=1,2$) are transverse components of the down-converted wave vectors.  The single photons with transverse wave vector $\boldsymbol{q}_l$ and frequency $\omega_l$ are written as  $\ket{\boldsymbol{q}_l}$.  The two-photon amplitude in the paraxial regime, written in wave vector coordinates at the exit face of the crystal is given by
\begin{equation}
\psi (\boldsymbol{q}_1,\boldsymbol{q}_2)= \mathcal{V}(\boldsymbol{q}_1+\boldsymbol{q}_2) \mathrm{sinc}[(k_{1 z}+k_{2 z}-k_{p z}) L / 2]. 
\label{eq:phipump}
\end{equation}
 Here $L$ is the length of the non-linear crystal and  $k_{p z}$ is the $z$-component of the pump beam wavevector.   The function $\mathcal{V}(\boldsymbol{q})$ is the angular spectrum of the pump beam \citep{monken98a}, and the $\mathrm{sinc}$ function is known as the phase matching function \citep{walborn10}.    For simplicity, all modes are assumed to be polarized.  If narrowband filters are used to detect the photons, we can assume monochromatic down-converted fields, and apply the Fresnel approximation $
k_z  \approx k (1-{q^2}/{2k^2})$, which gives 
\begin{align}
k_{1z}+k_{2z}-k_{pz}\approx  -\frac{1}{2 {k}_p} \left[\delta_1 \boldsymbol{q}_1- \delta_2  \boldsymbol{q}_2 \right]^2,
\end{align}
where we define $k_p \equiv \lvert\boldsymbol{k}_p\rvert$ as the wavenumber of the pump beam, and $\delta_1=\sqrt{k_2/k_1}$ , $\delta_2 =\sqrt{k_1/k_2}$. We can rewrite the two-photon amplitude as:
\begin{equation}
\psi (\boldsymbol{q}_1,\boldsymbol{q}_2)= \mathcal{V}(\boldsymbol{q}_1+\boldsymbol{q}_2) \mathcal{S} \left(\left[\delta_1 \boldsymbol{q}_1- \delta_2  \boldsymbol{q}_2 \right]^2 \right ),
\label{eq:phipump2}
\end{equation}
 where $\mathcal{S} ( \boldsymbol{q})=A \mathrm{sinc}(L q^2/ 4 k_p)$.
Defining the variables 
\begin{align}
\boldsymbol{Q}_{\pm} &=\boldsymbol{q}_1\pm\boldsymbol{q}_2, \label{Q+-}\\
{\boldsymbol{R}}_{\pm} &= \frac{1}{2} (\boldsymbol{r}_1 \pm \boldsymbol{r}_2) \label{eq:r+-} \\
\delta_{\pm} & = \frac{1}{2} ( \delta_1 \pm \delta_2 ), 
\end{align}
and taking the Fourier transform of  Eq. \eqref{eq:phipump2},  the two-photon amplitude can be written in position coordinates $\boldsymbol{r}_1,\boldsymbol{r}_2$ as 
\begin{align}
\Psi (\boldsymbol{r}_1,\boldsymbol{r}_2) = & \frac{1}{2} \iint  d \boldsymbol{Q}_+ d \boldsymbol{Q}_-  e^{i \boldsymbol{Q}_+ \boldsymbol{R}_+} e^{i \boldsymbol{Q}_- \boldsymbol{R}_-} \mathcal{V}(\boldsymbol{Q}_+) \nonumber \\ 
& \times 
  \mathcal{S} \left(\left[\delta_+ \boldsymbol{Q}_+ + \delta_- \boldsymbol{Q}_-\right]^2 \right ).
\label{eq:psipump1}
\end{align}
{By integrating} Eq.\eqref{eq:psipump1}, we obtain 
\begin{equation}
\Psi (\boldsymbol{r}_1,\boldsymbol{r}_2)= \frac{1}{2} \mathcal{W}\left (\boldsymbol{R}_+ - \frac{\delta_+}{\delta_-}\boldsymbol{R}_-\right) \Gamma \left (\boldsymbol{R}_-\right),
\label{eq:psipump2}
\end{equation}
where $\mathcal{W}$ and $\Gamma$ are the Fourier transforms of {the angular spectrum of the pump beam $\mathcal{V(\boldsymbol{q})}$ and the phase matching function $\mathcal{S(\boldsymbol{q})}$, respectively.} 

%%%%%%%%%%%%%%%%%%
\subsection{Projection onto PV modes}
%%%%%%%%%%%%%%%%%%
The amplitude to project the two-photon state onto a product state of PV modes is given by
\begin{align}
\mathcal{A}(\ell_1,\ell_2) =  \frac{1}{2} \iint & d\boldsymbol{r}_1 d\boldsymbol{r}_2  \Gamma \left (\frac{\boldsymbol{r}_1-\boldsymbol{r}_2}{2}\right) \mathcal{U}^*_{\ell_1}(\boldsymbol{r}_1)  \mathcal{U}^*_{\ell_2}(\boldsymbol{r}_2)\times \nonumber \\
& \mathcal{W}\left (\frac{\boldsymbol{r}_1+\boldsymbol{r}_2}{2} - \frac{\delta_+}{\delta_-}\frac{\boldsymbol{r}_1-\boldsymbol{r}_2}{2}\right). 
\label{eq:overlap}
\end{align}
In the thin crystal approximation, such that $L << z_R$, where $z_R$ is the Rayleigh range of the pump beam, we can approximate $\Gamma \left(\frac{\boldsymbol{r}_1-\boldsymbol{r}_2}{2}\right) \approx 2 \delta \left(\boldsymbol{r}_1-\boldsymbol{r}_2\right)$, where $\delta(x)$ is the Dirac delta function. 
The amplitude becomes
\begin{equation}
\mathcal{A}(\ell_1,\ell_2) = \int  d\boldsymbol{r}\mathcal{U}^*_{\ell_1}(\boldsymbol{r})  \mathcal{U}^*_{\ell_2}(\boldsymbol{r})
 \mathcal{W}\left (\boldsymbol{r}\right). 
\label{eq:overlap2}
\end{equation}
Assuming now that the pump beam is an OAM eigenstate, it can be written as $ \mathcal{W}\left (\boldsymbol{r}\right) = M \exp(i \ell \phi) w_\ell(r)$, with $M$ a normalization constant. Then,  using expression \eqref{eq:pvb} for the PV modes, we have
\begin{align}
\mathcal{A}(\ell_1,\ell_2,\ell) =   & (N^{*})^2 M \int  dr \, r u_{\ell_1}^*(r) u_{\ell_2}^*(r) w_\ell(r) \nonumber \\ 
& \times 
 \int d \phi e^{i \phi(\ell-\ell_1-\ell_2)},
\label{eq:overlap3}
\end{align}
which leads to 
\begin{equation}
\mathcal{A}(\ell_1,\ell_2,\ell) =  \delta_{\ell,\ell_1+\ell_2} (N^{*})^2 M' \int  dr \, r u_{\ell_1}^*(r) u_{\ell_2}^*(r) w_\ell(r),
\label{eq:overlap4}
\end{equation}
{where $M'=2\pi M$}.
{The appearance of the Kronecker delta function guarantees that the OAM winding numbers of the down-converted photons are correlated. These OAM correlations are typically observed in OAM mode decompositions of the two photon state, and corresponds to conservation of the orbital angular momentum} \citep{mair01,arnaut01,franke-arnold02,torres03b,walborn04a,law04,mclaren12,mclaren13,straupe11}.
%%%%%%%%%%%%%%%%%%
\subsection{Limited OAM spectra: optimizing into RCF modes}
%%%%%%%%%%%%%%%%%%

%%%%%%%%%%%%%%%%%%%
\begin{figure}
\begin{center}
\includegraphics[width=8cm]{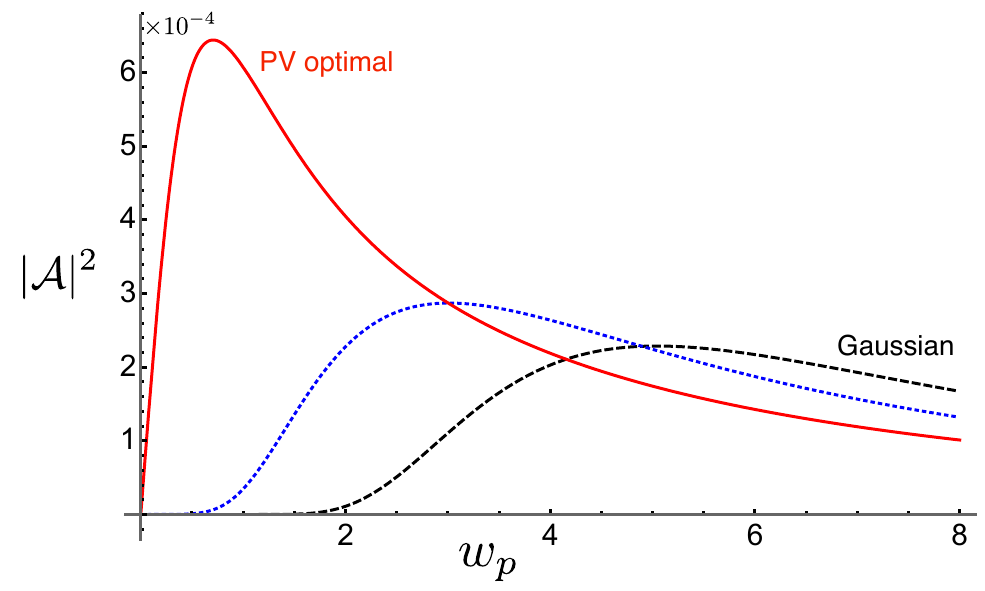}
\caption{Overlap squared $|\mathcal{A}|^2$ as a function of the pump beam width $w_p$ ({in arbitrary units of length)} for a gaussian pump beam $r_{rp}=0$ (black dashed curve), and for a PV pump beam with  ring radius $r_{rp}=2 w_0 \approx 0.57 r_r$ (blue dotted curve) and ring radius $r_{rp}=r_r=3.53w_0$ (PV optimal, red solid curve) for best coupling into  a commercially available RCF \citep{rojas21}. The optimal PV pump beam increases the probability to produce relevant down-converted modes by a factor of $\sim 2.8$.} 
\label{fig:plots}
\end{center}
\end{figure}
%%%%%%%%%%%%%%%%%%%
\label{ref:limspec}
Let us consider now {that the} OAM spectra of the down-converted photons {are} limited by the optical system, {such as is the case} when coupling into a RCF, which supports a finite set $\{L\}$ of OAM eigenmodes. We assume that $\ell_j \in \{L\}$ and the ratio $r_r/w_0$ of the down-converted {modes permits} approximation of $u_\ell(r) \approx u(r)$, where we recall that this is valid when $\ell_j \lesssim r_r/w_0$.  Then, we can write the PV mode product
\begin{equation}
u_{\ell_1}^*(r) u_{\ell_2}^*(r) \approx [u^*(r,r_r,w_0)]^2 =  u^*(r,r_r,w_0/\sqrt{2}),
\label{eq:u2}
\end{equation}
where the RCF ring radius and beam width are included explicitly in the argument of these functions for clarity.  The integral \eqref{eq:overlap4} is thus the overlap of a PV radial mode function described by a gaussian ring centered at $r_r$ and ring thickness $w_0/\sqrt{2}$, with the pump radial mode function $w_\ell(r)$.  It is thus expected that the amplitude integral is maximized when the pump beam is also a PV beam, with the same ring thickness and ring width.  Let us also consider that the pump is prepared in a PV mode and that $w_\ell(r) \approx u(r) \equiv u(r,r_{rp},w_p)$. The amplitude  integral \eqref{eq:overlap4} becomes
\begin{equation}
\mathcal{A} =   (N^{*})^2 M' \int  dr \, r\, u^*(r,r_r,w_0/\sqrt{2}) u(r,r_{rp},w_p),
\label{eq:overlap5}
\end{equation} 
which has the very appealing characteristic that it {depends neither on} $\ell$ nor $\ell_1, \ell_2$. In figure \ref{fig:plots} we show a plot of  $|\mathcal{A}|^2$ as a function of the pump beam width $w_p$ for several pump beams, where we used $r_r= 3.53 w_0$, corresponding to the parameters of PV modes that are most efficiently coupled into a commercially available RCF \citep{rojas21}.   The red solid curve and the blue dotted curve correspond to PV pump beams with ring radii $r_{rp}=r_r$ and $r_{rp}=2 w_0\approx 0.57 r_r$.  As a comparison, we also show the {squared} amplitude when the pump beam is described by a gaussian beam $w(r)=\sqrt{{2}/{(\pi  w_p^2)}} \exp(-{r^2}/{w_p^2})$.   We can see that the amplitude to produce down-converted photons in these eigenmodes can be increased by using a PV pump beam, and reaches a maximum when $w_p=w_0/\sqrt{2}$.  In principle, this corresponds to an a factor of $\sim 2.8$ increase in generation of the relevant down-converted modes.    We note that we have considered nearly co-linear SPDC, in which the pump beam and photon pair propagate in the same direction.  It has been shown that non-colinear SPDC can lead to asymmetry when the pump beam is a PV mode \citep{jabir16}. Recently, it was shown that the heralding efficiency of a twisted down-converted photon with $\ell_2=\ell$ (with $\ell_1=0$) can be increased by pumping with a PV mode \cite{anwar2021size}.    

\par
An important characteristic here is that, aside from the OAM correlation provided by the Kronecker delta function, the amplitude of the overlap coefficients does not depend upon $\ell_1$, $\ell_2$ nor $\ell$.  Thus, considering a post-selected set of PV modes in both the down-converted fields, the two-photon state can be written
\begin{equation}
\ket{\psi} =  \frac{\mathcal{A}}{\sqrt{C}} \sum_{\ell-\ell_1, \ell_1 \in \{ L \}} \ket{\ell_1}_1 \ket{\ell-\ell_1}_2,
\label{eq:state}
\end{equation}
where the states $\ket{\ell}$ represent single photon PV modes, i.e. $\langle \boldsymbol{r}\ket{\ell}=\mathcal{U}_{\ell}(\boldsymbol{r})$, {$C$ is a normalization constant,} and $\{L\}$ is the set of OAM supported by the fibers.  This is a maximally entangled state, whose entanglement depends upon the number of terms in the summation. We emphasize here the fact that \eqref{eq:state} is valid in the case where the  characteristics of the pump and down-converted photons allow the modes to be written in form \eqref{eq:pvb2}. Typically, this approach is restricted to smaller OAM values.  In the next section, we consider the complete two-photon OAM spectrum. 
%%%%%%%%%%%%%%%%%%
 \subsection{Unlimited OAM Spectra}
%%%%%%%%%%%%%%%%%% 
%%%%%%%%%%%%%%%%%%%
\begin{figure}
\begin{center}
\includegraphics[width=8cm]{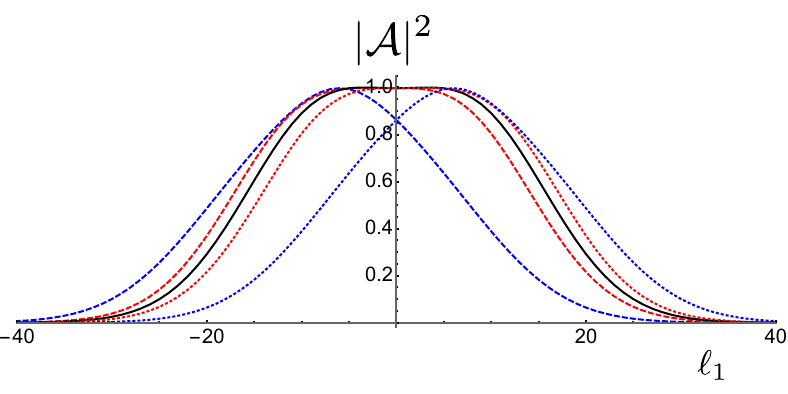}
\caption{Mode probabilities $|\mathcal{A}(\ell_1,\ell-\ell_1,\ell)|^2$, normalized by $|\mathcal{A}(0,0,0)|^2$, as a function of $\ell_1$ for a PV pump beams with optimal width $w_p=w_0/\sqrt{2}$ and ring radius $r_{rp}=r_r$, and OAM number $\ell=0$ (black solid line), $\ell=\pm 3$ (red dashed and dotted lines) and $\ell=\pm 12$ (blue dashed and dotted lines).  The curves are centered at $\ell_1 = \ell$.} 
\label{fig:probs}
\end{center}
\end{figure}
%%%%%%%%%%%%%%%%%%%
%%%%%%%%%%%%%%%%%%%
\begin{figure}
\begin{center}
\includegraphics[width=7cm]{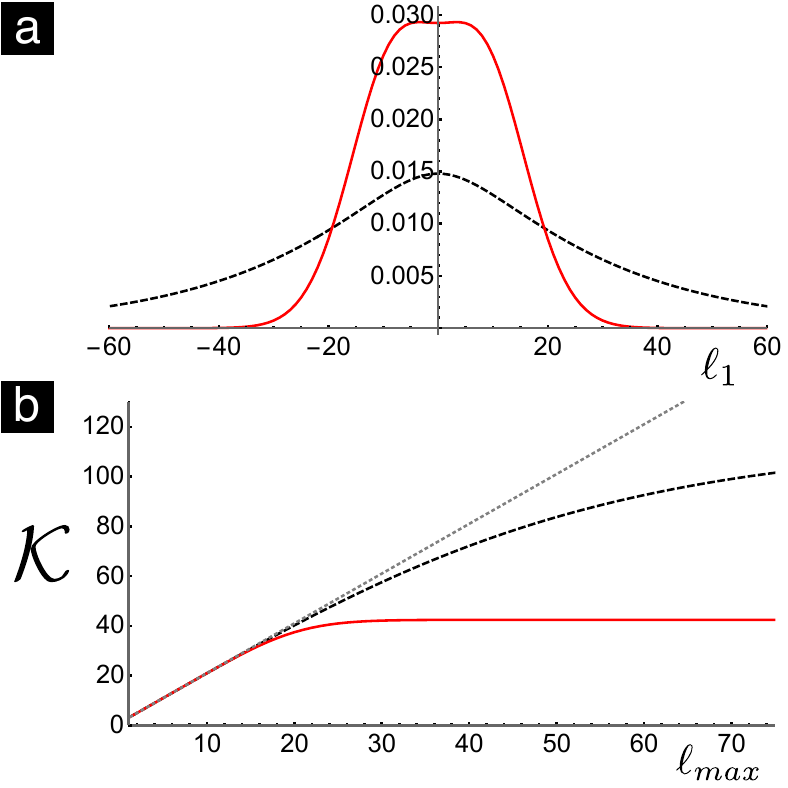}
\caption{a) Normalized probabilities for an optimal PV pump beam with $\ell=0$ (red solid line) and a gaussian pump beam with $w_p=5 w_0$ (black dashed line). The PV pump mode concentrates the probability in lower OAM modes. b) The Schmidt number $\mathcal{K}$ for the two photon state, truncated at $\pm \ell_{max}$ for the same pump beams. The grey dotted line corresponds to the maximum allowable Schmidt number, given by $d=2 \ell_{max}+1$.  } 
\label{fig:pvb-gauss}
\end{center}
\end{figure}
%%%%%%%%%%%%%%%%%%%
When the optical system does not severely limit the OAM spectra of the down-converted photons, such as emission into free space, we can evaluate the amplitude integral  \eqref{eq:overlap4} numerically using Eq. \eqref{eq:pvb3} to describe the modes.  Plots of $|\mathcal{A}(\ell_1,\ell-\ell_1,\ell)|^2$ as a function of $\ell_1$ are shown in Fig. \ref{fig:probs} for $\ell=0,\pm3,\pm12$.  To compare the relative weights, we have normalized to the value $|\mathcal{A}(0,0,0)|^2$. We can see that the curves have similar {forms and magnitudes}, but are shifted along the $\ell_1$ axis by an amount equal to $\ell$.  

\par
To compare between a PV and a gaussian pump beam, we evaluated the OAM mode spectra and the entanglement of the two-photon state. Figure \ref{fig:pvb-gauss} a) shows the normalized probabilites $|\mathcal{A}_{\ell_1}|^2/\sum_{\ell_1} |\mathcal{A}_{\ell_1}|^2$ for a PV pump beam given by \eqref{eq:pvb3} with the optimal parameters $w_p=w/\sqrt{2}$ and $r_{rp}=r_r= 3.53 w_0$, and a gaussian beam with $w_p=5 w$, similar to section \ref{ref:limspec}.  We can see that the probability to produce down-converted PV modes in the range $|\ell_{j}| \lesssim 15\, (j=1,2)$ is larger for the PV pump beam than the gaussian pump beam, and nearly constant for $|\ell_{j}| \lesssim 6$.  Moreover, the probabilities approach zero for $|\ell_{j}| \sim 30$. The gaussian pump mode, on the other hand, results in an OAM spectrum with much more spread, giving negligible probabilities only when $|\ell_{j}| \gtrsim 100$ (not shown).  {The narrower mode spectrum produced by the PV pump beam concentrates the probability in a smaller group of OAM modes, leading to better efficiency when one is working within this finite subspace. For example, looking at the subset of modes corresponding to $|\ell| \leq 15$ (a $31 \times  31$ dimensional bipartite system) in Fig. \ref{fig:pvb-gauss} a), this corresponds to 81$\%$ of the state produced with the PV pump beam, and only 41$\%$ percent of the state produced by the gaussian pump beam.   Likewise, for generating the 13 OAM modes ($|\ell| \leq 6$) of the RCF fiber studied in Ref. \cite{rojas21}, the PV pump beam concentrates 38$\%$ of the probability in these modes, compared to 19$\%$ for the gaussian pump beam.} 
\par
It is known that SPDC can produce high-dimensional entanglement in transverse spatial modes \citep{walborn07c,bavaresco18,erhard18,schneeloch19,forbes19,erhard20}.  Let us see how the PV pump beam compares with a gaussian pump beam for high-dimensional entanglement generation. 
The entanglement in PV modes can be evaluated by calculating the Schmidt number, given by
\begin{equation}
\mathcal{K} = \frac{\left( \sum\limits^{\ell_{max}}_{\ell_1=-\ell_{max}} \mathcal{A}_{\ell_1}\right)^2}{\sum\limits^{\ell_{max}}_{\ell_1=-\ell_{max}} \mathcal{A}_{\ell_1}^2},
\end{equation}
where we assume that the relevant OAM spectrum is limited by $\pm \ell_{max}$, so that the overall dimension of the system is $d=2  \ell_{max} + 1$. 
In Fig.   \ref{fig:pvb-gauss} b) we calculate $\mathcal{K}$ for different values of $\ell_{max}$ for the same pump beams as in   \ref{fig:pvb-gauss} a).  The dotted grey line shows the maximum allowable $\mathcal{K}=d$ for comparison. We can see that the PV pump beam results in a Schmidt number that saturates at around $\mathcal{K} \sim 42$, while the gaussian pump beam saturates at  $\mathcal{K} \sim 128$ for $\ell_{max} \sim 150$ (not shown).   For smaller values of $\ell_{max}$, the PV beam and gaussian beam give almost equivalent Schmidt number, and nearly saturate the maximum allowed value of $d$.  We also calculated the largest Schmidt numbers for PV pump modes with $|\ell| \leq 12$, and obtained nearly constant results in the range $\mathcal{K} \sim 40-43$.  Thus, if the optical system imposes no practical limit on the system dimension, the Gaussian beam allows for higher-dimensional entanglement. Nonetheless, for the more realistic scenario in which the optical system  supports a finite set of modes with smaller values of $|\ell_j|$, a PV pump beam concentrates the probability into a smaller set of modes, resulting in {a more efficient} source, and near maximal entanglement.

%%%%%%%%%%%%%%%%%%%
\section{Application to Quantum State Engineering}
\label{Sec:Apps}
%%%%%%%%%%%%%%%%%%%
%%%%%%%%%%%%%%%%%%%
\begin{figure}
\begin{center}
\includegraphics[width=7cm]{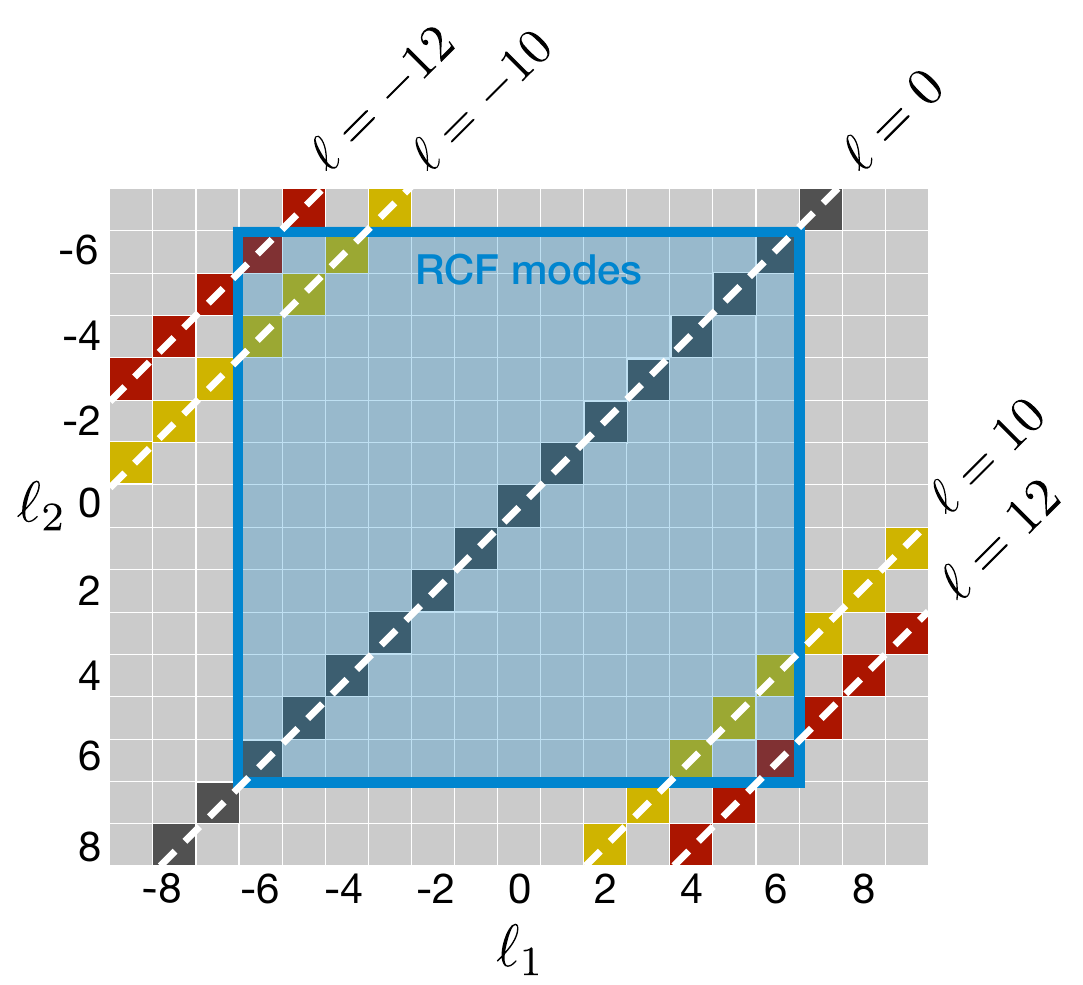}
\caption{Illustration of state engineering using OAM correlations from SPDC and mode filtering provided by ring core fibers (RCF). OAM correlations between the pump beam and down-converted photons produce photons with OAM distributed along the diagonal directions. The RCF selects modes within the shaded blue square.  Thus, only OAM mode pairs along the diagonals within the square will propagate.} 
\label{fig:app}
\end{center}
\end{figure}
%%%%%%%%%%%%%%%%%%%
 Engineering of quantum states with different properties is both a challenge and a goal in quantum information science.  In SPDC, this can be achieved by manipulation of the properties of the pump beam, as well as through mode filtering of the individual photons. In this regard, integrating the SPDC results from the last sections together with the transmission properties of RCFs presents several interesting possibilities. 
\par
As shown in Figs. \ref{fig:probs} and \ref{fig:pvb-gauss}, the PV pump mode concentrates the two-photon probability into a smaller set of joint OAM modes, which is particularly interesting when the photons are coupled into devices that support a finite number of eigen-modes. 
Let us consider that both down-converted photons are coupled to RCFs that support OAM eigenmodes with $|\ell_1|, |\ell_2|  \leq 6$, as was studied in Ref. \citep{rojas21}. Thus, each down-converted photon has an OAM spectrum that is truncated in the shaded blue square region shown in Fig. \ref{fig:app}.  When the pump beam is described by a single PV mode with OAM $\ell$, the two-photon state is described by \eqref{eq:state}.  The important point here is that the state \eqref{eq:state} contains joint OAM modes with $\ell_2=\ell-\ell_1$, non-zero overlap integral \eqref{eq:overlap5}, and with $|\ell_1|, |\ell_2|  \leq 6$. These three conditions can be used to engineer the quantum state. Fig. \ref{fig:app} illustrates the allowable joint OAM spectra for different pump beams, where allowed mode combinations appear on the diagonals, as a function of the pump OAM number $\ell$, and within the blue square region, corresponding to the mode selection of the RCF. Since for this set of modes the overlap integral is approximately given by \eqref{eq:overlap5}, which is independent of the the OAM of the pump and down-converted fields, the pump OAM $\ell$ can be used as a parameter to control the entanglement, where the Schmidt number of the state is essentially determined by the number of OAM components distributed along diagonals and within the RCF square.  Thus, one can achieve a 13-dimensional entangled state by pumping with $\ell=0$. Alternatively,  a separable product state can be achieved by pumping with $\ell=\pm 12$, which gives a two-photon state $\ket{\psi}= \ket{\pm 6}_1\ket{\pm 6}_2$, since these are the only joint OAM modes that are both produced in SPDC and supported by the RCF. 

\par
A maximally entangled pair of qubits is arguably the most useful quantum state, with numerous applications in quantum information, such as teleportation and quantum key distribution.  This state can be created by using a pump beam that is a superposition of PV modes.    For example, using a pump beam described by a PV mode superposition $\mathcal{E}(\boldsymbol{r})= \alpha \mathcal{U}_{12} (\boldsymbol{r})+\beta \mathcal{U}_{-12} (\boldsymbol{r})$, with intensity profile illustrated in Fig. \ref{fig:rcf} d), the two-photon state that propagates in the RCFs modes is 
 \begin{equation}
\ket{\psi_{\pm12}} =  \alpha \ket{6}_1 \ket{6}_2 + \beta \ket{-6}_1 \ket{-6}_2.
\label{eq:state3}
\end{equation}
Maximal entanglement is achieved when $|\alpha|=|\beta|$.  Similar Bell-type states have been prepared in OAM modes from SPDC, however, they rely on post-selection at the detection system \citep{mair01,langford04}. 
Moreover, we note that here the OAM numbers are correlated ($\ell_1=\ell_2$), as opposed to anti-correlated ($\ell_1=-\ell_2$), as is usually the case due to OAM conservation. Correlated OAM states have been shown to be useful for quantum metrology \citep{dambrosio13b}.
\par
By the same rationale, pumping with an equally weighted superposition of $\pm 10$, we have
 \begin{align}
\ket{\psi_{\pm10}} =   & \frac{1}{\sqrt{6}}( \ket{4}_1 \ket{6}_2 + \ket{5}_1 \ket{5}_2  
 +  \ket{6}_1\ket{4}_2 + \nonumber \\  &  \ket{-4}_1 \ket{-6}_2 +  \ket{-5}_1 \ket{-5}_2  
+  \ket{-6}_1\ket{-4}_2),
\label{eq:state4}
\end{align}
which is a $6\times6$ maximally entangled state.  Higher-dimensional states can provide higher key transmission rates in quantum key distribution \citep{bourennane01,collins02,cerf02,walborn06a,huber13,Canas17}, as well as increased resilience to noise \citep{zhu21} and other applications \cite{erhard18,forbes19,erhard20}. 
These are just a few simple examples of how the dimension and entanglement of the two-photon state can be controlled by manipulating the PV pump beam {and post-selection capabilities of the ring-core fiber.} 
More complex quantum states can be created by considering different linear combinations of PV pump beams.

\section{Conclusion}
The generation of entangled photons in perfect vortex modes was studied in the spontaneous parametric down-conversion process. Perfect vortex modes carry orbital angular momentum, and have very high fidelity with the eigen-modes of ring-core optical fibers. We show that pumping the non-linear crystal with a perfect-vortex beam, leads to an output two-photon state that is concentrated in a smaller set of modes, when compared to that of a Gaussian pump beam.  It is shown that a near three-fold  increase in the coupling efficiency into ring-core fibers could be achieved.  Moreover, the two-photon mode spectrum can have near constant magnitude, allowing for a high degree of entanglement.
\par
The use of ring-core fibers as mode filters together with pump beam engineering can be a powerful tool for crafting novel quantum states.  Several examples are given, ranging from product states to $13 \times 13$ dimensional entangled states. These can be produced by changing a single pump beam parameter. Though we focus only on OAM modes, our findings can be combined with correlations in other degrees of freedom, such as polarization.  We expect our results will be important for integrating entangled photon sources with future optical fiber networks that employ structured optical fibers.

\section*{Funding}
 This work was supported by the Chilean agencies Fondo Nacional de Desarrollo Cient\'{i}fico y Tecnol\'{o}gico (FONDECYT) (1190933, 1190901, 1200266, 1200859) and ANID – Millennium Science Initiative Program – ICN17\_012.

\section*{Acknowledgments}
The authors thank S. E. Restrepo for valuable discussions.

% Please see the availability of data guidelines for more information, at https://www.frontiersin.org/about/author-guidelines#AvailabilityofData

%\bibliographystyle{apsrev}
%\bibliographystyle{frontiersinHLTH&FPHY}
%\bibliographystyle{frontiersinSCNS_ENG_HUMS} % for Science, Engineering and Humanities and Social Sciences articles, for Humanities and Social Sciences articles please include page numbers in the in-text citations
%\bibliographystyle{frontiersinHLTH&FPHY} % for Health, Physics and Mathematics articles

%\bibliography{Master_Bibtex-SPDCPVB}

\end{document}